\documentclass{svproc}

\usepackage[USenglish]{babel}
\usepackage[utf8]{inputenc}
\usepackage[T1]{fontenc}
\usepackage{graphicx}
\usepackage{amsmath}
\usepackage{amssymb}
\usepackage{bm}
\usepackage{xcolor}
\usepackage{import}
\usepackage{multirow}
\usepackage[sectionbib,numbers]{natbib} 

\newcommand{\mbf}[1]{\ensuremath{{\bm{#1}}}}

\newcommand{\JacobianoLinear}{\bm{J}}
\newcommand{\JacobianoAngular}{\bm{W}}

\newcommand{\FrameC}[1]{\ensuremath{\mathcal{C}_{#1}}}
\newcommand{\InertialFrame}{\ensuremath{\mathcal{I}}}
\newcommand{\BodyFrame}{\ensuremath{\mathcal{B}}}
\newcommand{\RotationMatrixInertialBody}{\ensuremath{{\bm{R}^\InertialFrame_\BodyFrame}}}
\newcommand{\RotationMatrix}[2]{{\mbf{R}^{{#1}}_{{#2}}}}
\newcommand{\PositionVector}[3]{\mbf{d}_{{#2},{#3}}^{{#1}}}
\newcommand{\PPositionVector}[3]{\mbf{p}_{{#2},{#3}}^{{#1}}}
\newcommand{\dPPositionVector}[3]{\dot{\mbf{p}}_{{#2},{#3}}^{{#1}}}
\newcommand{\OnesMatrix}[2]{\ensuremath{\mbf{1}}}
\newcommand{\ZerosMatrix}[2]{\ensuremath{\mbf{0}}}
\newcommand{\IdentityMatrix}[1]{\mbf{I}}

\newcommand{\SkewSymmetricMatrix}[1]{{\mbf{S}\big({{#1}}\big)}}
\newcommand{\Propeller}[1]{\bm{\Omega}_{{#1}}}
\newcommand{\PPropeller}[1]{\mathcal{P}_{{#1}}}
\newcommand{\VelocidadeAngular}[3]{\mbf{w}^{{#1}}_{{#2},{#3}}}
\newcommand{\VelocidadeLinear}[3]{\mbf{v}^{{#1}}_{{#2},{#3}}}

\newcommand{\realset}{\ensuremath{\mathbb{R}}}
\newcommand{\realsetmat}[2]{\ensuremath{\mathbb{R}^{#1\times#2}}}

\newcommand{\eye}[1]{\ensuremath{\mbf{I}_{#1}}}

\newcommand{\zeros}[2]{\ensuremath{\mbf{0}_{#1\times#2}}}
\newcommand{\ones}[2]{\ensuremath{\mbf{1}_{#1\times#2}}}

\newcommand{\pderiv}[2]{\ensuremath{\partial_{{#2}}{#1}}}
\newcommand{\pderivtrans}[2]{\ensuremath{\partial'_{{#2}}{#1}}}
\newcommand{\pderivsecond}[2]{\ensuremath{\partial^2_{{#2}}{#1}}}

\newcommand{\half}{\ensuremath{0.5}}

\definecolor{orange}{RGB}{255,69,0}

\allowdisplaybreaks

\usepackage{url}

\begin{document}
\mainmatter              
\title{Joint State-Parameter Observer-Based Robust Control of a UAV for Heavy Load Transportation\thanks{This work was in part supported by the project INCT (National Institute of Science and Technology) for Cooperative Autonomous Systems Applied to Security and Environment under the grants CNPq 465755/2014-3 and FAPESP 2014/50851-0, and by the Brazilian agencies CAPES under the grant numbers 88887.136349/2017-00 and 001, CNPq under the grant 315695/2020-0, FAPEMIG under the grant APQ-03090-17, and FAPESP under the grant 2022/05052-8.}}
\titlerunning{State-Parameter Observer-Based Control of a UAV for Load Transportation}  
%
\author{Brenner S. Rego\inst{1} \and Daniel N. Cardoso\inst{2} \and Marco. H. Terra\inst{1} \and Guilherme V. Raffo\inst{2}}

\authorrunning{Rego et al.} 
%
\tocauthor{Brenner S. Rego, Daniel N. Cardoso, Marco. H. Terra, Guilherme V. Raffo}
\institute{University of São Paulo, São Carlos, SP 13566-590, Brazil,\\
\email{brennersr7@usp.br,terra@sc.usp.br},
\and
Federal University of Minas Gerais,
Belo Horizonte, MG 31270-901, Brazil,\\
\email{\{danielneri,raffo\}@ufmg.br}}

\maketitle              
\vspace{-1mm}
\begin{abstract}
\vspace{-6mm}
This paper proposes a joint state-parameter observer-based controller for trajectory tracking of an octocopter unmanned aerial vehicle (OUAV), for transportation of a heavy load with unknown mass and size. The multi-body dynamic model of the OUAV with a rigidly attached load is obtained, effectively considering the effects of the load parameters into the dynamics of the system. A robust nonlinear $\mathcal{W}_\infty$ control strategy is designed for optimal trajectory tracking of the OUAV, with information of the states and load parameters provided by a joint estimation unscented Kalman filter. The effectiveness of the proposed strategy is corroborated by numerical results.
\vspace{-1mm}
\end{abstract}

\vspace{-8mm}
\section{Introduction}

Taking advantage of their versatility and autonomous operation, unmanned aerial vehicles (UAVs) can be used for aerial load transportation, with many applications such as vertical replenishment of seaborne vessels \citep{Wang2014}, deployment of supplies in search-and-rescue missions \citep{Bernard2011}, package delivery, and landmine detection \citep{BisgaardThesis}. Aerial load transportation using UAVs is a challenging task in terms of modeling and control. The load may be connected to the UAV either rigidly or by means of a rope, which changes its dynamics considerably. In addition, the load physical parameters are often unknown in practice, and their knowledge is usually necessary to effectively accomplish the task. %
A model-free control approach based on trajectory generation by reinforcement learning has been proposed in \cite{Palunko2013} for path tracking of the load using a quadrotor UAV (QUAV). However, this is an open-loop method, and the entire learning process must be repeated if there are any changes in the system. A hybrid control strategy has been proposed in \cite{Sreenath2013b} addressing a similar transportation problem, relying on a decoupling-based cascade structure, and considering the case in which the cable is taut and when it is loose. However, the strategy is not robust to external disturbances and unmodeled dynamics. 

Although several works have investigated the load transportation problem, only a few propose estimation techniques to obtain information on the load physical parameters, which are important to achieve effectiveness in the task. An unscented Kalman filter (UKF) has been designed in \cite{Bisgaard2010} to estimate the load's position and velocity for a helicopter with suspended load. The authors have also proposed methods for estimating the wire's length. This algorithm has been combined with delayed adaptive feedback control, taking advantage of the wire's length estimation. Nevertheless, other physical parameters such as the load mass, are assumed to be known. State and parameter estimation has been addressed in \cite{Prkacin2020} for load transportation using a QUAV. The proposed methodology estimates the angles and length of the cable; however, it assumes the load mass to be precisely known, in addition to disregard the load inertia tensor, being unsuitable for heavy load transportation. Adaptive control has been investigated in \cite{Yu2022} considering the estimation of the load mass, but the rotational dynamics of the load have been once again neglected. Lastly, a geometric control strategy based on an inner-outer loop structure with disturbance estimation has been proposed in \cite{Gajbhiye2022}. Nevertheless, it has used a simplified model where the inertia tensor of the load has been again neglected.

In view of these challenges, we propose a joint state-parameter observer-based controller for trajectory tracking of an octocopter UAV (OUAV) that carries a heavy load of unknown mass and size. The contributions are threefold: (i) the multi-body dynamic modeling of the OUAV with a rigidly attached heavy load; (ii) the design of a novel robust nonlinear control strategy with fast disturbances attenuation for load transportation of the OUAV, based on the proposal of a new particular solution to the nonlinear $\mathcal{W}_\infty$ control problem \cite{AUT2019}; and (iii) the design of a UKF-based algorithm for robust joint estimation of the OUAV states, exogenous disturbances, and the load parameters, which are provided to the nonlinear $\mathcal{W}_\infty$ controller for efficient trajectory tracking of the OUAV with load.

\section{Octocopter UAV Modeling}
\label{OctocopterUAVmodel}

\begin{figure}[!htb]
	\centering{
	\def\svgwidth{0.98\columnwidth}
	{\tiny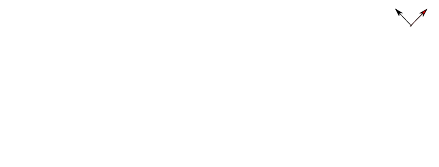}
    \caption{Definitions: propeller velocities $\Propeller{p}$ and frames $\PPropeller{p}$, COMs frames $\FrameC{O}$, $\FrameC{L}$, inertial frame $\InertialFrame$, body frame $\BodyFrame$, and angles $\alpha_p$, with $p \in \{1,\ldots,8\}$.}\label{QR}}
\end{figure}

In this work, the UAV equations of motion are obtained using the Euler-Lagrange formalism (ELF) \citep{spong2006robot}, considering the UAV and the load as two rigidly attached bodies. We define four reference frames (see Figure \ref{QR}): the inertial frame $\InertialFrame$, the body frame $\BodyFrame$, which is rigidly attached to the OUAV, and the frames $\FrameC{O}$ and $\FrameC{L}$ that are rigidly attached at the centers of mass (COMs) of the OUAV and load, respectively.

The OUAV is assumed to have six degrees of freedom (DOF), being the generalized coordinates vector defined as \small$\bm{q}(t) \triangleq \begin{bmatrix} \bm{\xi}' & \bm{\eta}' \end{bmatrix}'$\normalsize, where \small$\bm{\eta}(t) \triangleq \begin{bmatrix} \phi & \theta & \psi \end{bmatrix}'$\normalsize, is the orientation of $\BodyFrame$ w.r.t. $\InertialFrame$ described by the Euler angles using the $ZYX$ convention about the local axes, with \small$\phi(t), \theta(t), \psi(t) \in \mathbb{R}$\normalsize; and \small$\bm{\xi}(t) \triangleq \begin{bmatrix} x & y & z \end{bmatrix}'$ \normalsize is the position of the origin of frame $\BodyFrame$ w.r.t. $\InertialFrame$, with \small$x(t), y(t), z(t) \in \mathbb{R}$\normalsize.
Accordingly, the orientation and position of the origin of $\BodyFrame$ w.r.t. $\InertialFrame$ are given, respectively, by\footnote{Throughout the manuscript, some function dependencies are omitted. Moreover, $\IdentityMatrix{m}$ and $\ZerosMatrix{m}{m}$ are identity and zero matrices, respectively, with appropriate dimensions; $||\bm{z}(t)||_{\mathcal{W}_{\kappa,p,\bm{\Gamma}}}  \triangleq \big(\textstyle\sum_{\alpha = 0}^{\kappa} ||{d^{\alpha}\bm{z}(t)}/{dt^{\alpha}}||_{\mathcal{L}_p,\bm{\Gamma}_\alpha}^p\big)^{{1}/{p}}$, with $\bm{\Gamma}  \triangleq \{\bm{\Gamma}_0, ..., \bm{\Gamma}_\kappa\}$,
where $p \in \mathbb{N}\cup\{\infty\}$, $\kappa \in \mathbb{N}\cup\{0\}$ and $||\bm{z}(t)||_{\mathcal{L}_p,\bm{\Gamma}_\alpha} \triangleq \big(\int_{0}^{\infty} ||\bm{\Gamma}_\alpha^{1/p}\bm{z}(t)||_p^p~dt\big)^{\frac{1}{p}}$, in which $\bm{\Gamma}_\alpha > 0$; and $\pderiv{V}{t} \triangleq \partial V /\partial t$, $\pderiv{V}{\bm{\chi}} \triangleq \partial V /\partial \bm{\chi}$.} \small$\RotationMatrix{\InertialFrame}{\BodyFrame} = \bm{R}_{z,\psi}\bm{R}_{y,\theta}\bm{R}_{x,\phi}$\normalsize, and \small$\PPositionVector{\InertialFrame}{\InertialFrame}{\BodyFrame} = \bm{\xi}$\normalsize, 
and the orientation and position of the COMs of the OUAV and load are  \small$\RotationMatrix{\InertialFrame}{i} = \RotationMatrix{\InertialFrame}{\BodyFrame}$\normalsize, and \small$\PPositionVector{\InertialFrame}{\InertialFrame}{i} = \PPositionVector{\InertialFrame}{\InertialFrame}{\BodyFrame} +  \RotationMatrix{\InertialFrame}{\BodyFrame}  \PositionVector{\BodyFrame}{\BodyFrame}{i}$\normalsize, for \small$i \in \{\FrameC{O}, \FrameC{L}\}$\normalsize, where \small$\PositionVector{\BodyFrame}{\BodyFrame}{i} \in \mathbb{R}^3$ \normalsize is the position of the origin of frame $i$ w.r.t. $\BodyFrame$.

The linear velocities of the COMs of the OUAV and load are computed through the time derivative of their respective positions 
\small\begin{align}
\textstyle
\dPPositionVector{\InertialFrame}{\InertialFrame}{i} = \VelocidadeLinear{\InertialFrame}{\InertialFrame}{i} 
= \dot{\bm{\xi}} +  \RotationMatrix{\InertialFrame}{\BodyFrame} \SkewSymmetricMatrix{\VelocidadeAngular{\BodyFrame}{\InertialFrame}{\BodyFrame}} \PositionVector{\BodyFrame}{\BodyFrame}{i}
= \begin{bmatrix} \IdentityMatrix{3} &  -  \RotationMatrix{\InertialFrame}{\BodyFrame} \SkewSymmetricMatrix{\PositionVector{\BodyFrame}{\BodyFrame}{i}}\JacobianoAngular_\eta
\end{bmatrix}\dot{\bm{q}}
= \JacobianoLinear_{i}\dot{\bm{q}},
\label{LinQuad} 
\end{align}\normalsize
where \small$\JacobianoLinear_{i}$ \normalsize is the linear velocity Jacobian, \small$\VelocidadeAngular{\BodyFrame}{\InertialFrame}{\BodyFrame} {\triangleq} \JacobianoAngular_\eta\dot{\bm{\eta}}$ \normalsize is the OUAV angular velocity vector, with \small$\VelocidadeAngular{\BodyFrame}{\InertialFrame}{\BodyFrame} \in \mathbb{R}^3$\normalsize, and \small$\JacobianoAngular_\eta {\triangleq} \begin{bmatrix}
		1 & 0 & {-}\sin(\theta) \\
		0 & \cos(\phi) & \cos(\theta)\sin(\phi) \\
		0 & {-}\sin(\phi) & \cos(\phi)\cos(\theta)
	\end{bmatrix}$ \normalsize
is the Euler matrix. Besides, we used the properties \small$\dot{\RotationMatrix{\InertialFrame}{\BodyFrame}} = \RotationMatrix{\InertialFrame}{\BodyFrame}\bm{S}(\VelocidadeAngular{\BodyFrame}{\InertialFrame}{\BodyFrame})$ \normalsize and \small$\SkewSymmetricMatrix{\VelocidadeAngular{\BodyFrame}{\InertialFrame}{\BodyFrame}} \PositionVector{\BodyFrame}{\BodyFrame}{i} = -\SkewSymmetricMatrix{\PositionVector{\BodyFrame}{\BodyFrame}{i}} \VelocidadeAngular{\BodyFrame}{\InertialFrame}{\BodyFrame}$\normalsize, where \small$\bm{S}(\cdot) \in \mathbb{R}^{3\times3}$ \normalsize is a skew-symmetric matrix \cite{spong2006robot}. In addition, the angular velocities of the COMs of the OUAV and load are given by
\small\begin{align}
\VelocidadeAngular{\InertialFrame}{\InertialFrame}{i} &= \RotationMatrix{\InertialFrame}{\BodyFrame} \VelocidadeAngular{\BodyFrame}{\InertialFrame}{\BodyFrame} 
= \RotationMatrixInertialBody \JacobianoAngular_\eta \dot{\bm{\eta}}
= \begin{bmatrix} \ZerosMatrix{3}{3} & \RotationMatrixInertialBody\JacobianoAngular_\eta\end{bmatrix}\dot{\bm{q}}
= \JacobianoAngular_{i}\dot{\bm{q}}, \quad \JacobianoAngular_{i} \triangleq \begin{bmatrix} \ZerosMatrix{3}{3} & \RotationMatrixInertialBody\JacobianoAngular_\eta\end{bmatrix}.
\label{AngQuad}
\end{align} \normalsize

Using the ELF \cite{spong2006robot}, the OUAV equations of motion are written as
\small\begin{equation}\label{EqCanonicaEulerLagrangeQuad}
\bm{M}(\bm{q})\ddot{\bm{q}}(t) + \bm{C}(\bm{q},\dot{\bm{q}})\dot{\bm{q}}(t) + \bm{g}(\bm{q}) = \bm{\vartheta}(\bm{q},\bm{\tau}) + \bm{\zeta}(t),
\end{equation} \normalsize
where \small$\bm{M}(\bm{q}) \in  \mathbb{R}^{6\times6}$ \normalsize is the inertia matrix, \small$\bm{C}(\bm{q},\dot{\bm{q}}) \in \mathbb{R}^{6\times6}$ \normalsize is the Coriolis and centripetal forces matrix, \small$\bm{g}(\bm{q}) \in \mathbb{R}^{6}$ \normalsize is the gravitational force vector, \small$\bm{\vartheta}(\bm{q},\bm{\tau}) \in \mathbb{R}^{6}$ \normalsize is the vector of generalized inputs, with \small$\bm{\tau}(t) \triangleq \begin{bmatrix} f_{\PPropeller{1}} & f_{\PPropeller{2}} & \cdots
& f_{\PPropeller{8}} \end{bmatrix}'$\normalsize, in which \small$f_{p}(t)$ \normalsize is the force applied by the $p$-th propeller, for \small$p \in \{1, \;2, ..., \; 8\}$\normalsize, and \small$\bm{\zeta}(t) \triangleq \begin{bmatrix} \bm{\zeta}_x & \bm{\zeta}_y & \bm{\zeta}_z & \bm{\zeta}_\phi & \bm{\zeta}_\theta & \bm{\zeta}_\psi\end{bmatrix}'$\normalsize, with \small$\bm{\zeta}(t) \in \mathbb{R}^{6}$\normalsize, is the vector of generalized disturbances.

Taking into account \eqref{LinQuad} and \eqref{AngQuad}, the inertia matrix is obtained from the system total kinetic energy \small$\mathcal{K}(\bm{q},\dot{\bm{q}}) = \sum_{i \in \{\FrameC{O}, \FrameC{L}\}}  \frac{1}{2} ( m_i ({\VelocidadeLinear{\InertialFrame}{\InertialFrame}{i}})'\VelocidadeLinear{\InertialFrame}{\InertialFrame}{i} +  ({\VelocidadeAngular{\InertialFrame}{\InertialFrame}{i}})' \RotationMatrix{\InertialFrame}{i}\bm{\mathbb{I}}_i\RotationMatrix{\InertialFrame}{i}' \VelocidadeAngular{\InertialFrame}{\InertialFrame}{i})$ $\triangleq \dfrac{1}{2}\dot{\bm{q}}'\bm{M}(\bm{q})\dot{\bm{q}} = \dfrac{1}{2}\dot{\bm{q}}'\begin{bmatrix}
\bm{M}_{11} & \bm{M}_{21}' \\
\bm{M}_{21} & \bm{M}_{22}
\end{bmatrix}\dot{\bm{q}}$\normalsize, 
where  \small$\bm{M}_{11} \triangleq (m_{\FrameC{O}} + m_{\FrameC{L}})\IdentityMatrix{3}$\normalsize, 
    \small$\bm{M}_{22} \triangleq \JacobianoAngular'_\eta \big( -m_{\FrameC{O}}\SkewSymmetricMatrix{\PositionVector{\BodyFrame}{\BodyFrame}{\FrameC{O}}}^2 - m_{\FrameC{L}}\SkewSymmetricMatrix{\PositionVector{\BodyFrame}{\BodyFrame}{\FrameC{L}}}^2\big)\JacobianoAngular_\eta + \JacobianoAngular'_\eta \big(\mathbb{I}_{\FrameC{O}} + \mathbb{I}_{\FrameC{L}}\big)\JacobianoAngular_\eta$\normalsize,
   \small$\bm{M}_{12} \triangleq  -\RotationMatrix{\InertialFrame}{\BodyFrame}\big(
    m_{\FrameC{O}}\SkewSymmetricMatrix{\PositionVector{\BodyFrame}{\BodyFrame}{\FrameC{O}}}+m_{\FrameC{L}}\SkewSymmetricMatrix{\PositionVector{\BodyFrame}{\BodyFrame}{\FrameC{L}}}
    \big)\JacobianoAngular_\eta$\normalsize,
    and \small$m_i \in \mathbb{R}$ and $\mathbb{I}_i \in \mathbb{R}^{3\times3}$ \normalsize  are, respectively, the mass and the inertia tensor matrix of the body whose $i$-th frame is rigid attached. 
The gravitational force vector is given by \small$\bm{g}(\bm{q}) = \pderiv{\mathcal{P}(\bm{q})}{\bm{q}}$\normalsize, where \small$\mathcal{P}(\bm{q}) = - \sum_{i \in \{\FrameC{O}, \FrameC{L}\}} m_{i} \bm{g_r}' \PPositionVector{\InertialFrame}{\InertialFrame}{i}$ \normalsize is the system total potential energy, and \small$\bm{g_r} \triangleq \begin{bmatrix} 0 & 0 & g \end{bmatrix}'$\normalsize, in which \small$g$ \normalsize is the gravitational acceleration, while \small$\mbf{C}(\mbf{q},\dot{\mbf{q}})$ \normalsize is obtained from \small$\mbf{M}(\mbf{q})$ \normalsize by computing the Christoffel symbols of first kind \citep{spong2006robot}.

The force and torque generated by the $p$-th propeller, \small$p \in \{1, 2, ..., 8\}$\normalsize, are assumed to be applied at its geometric center. Then, to compute \small$\bm{\vartheta}(\bm{q},\bm{\tau})$\normalsize, a frame is rigidly attached to the geometric center of the $p$-th propeller, denoted by \small$\PPropeller{p}$\normalsize, as illustrated in Figure \ref{QR}. Accordingly, the orientation and position of the origin of \small$\PPropeller{p}$ \normalsize w.r.t. $\InertialFrame$ are, respectively, \small$\RotationMatrix{\InertialFrame}{\PPropeller{p}}  =  \RotationMatrix{\InertialFrame}{\BodyFrame}$\normalsize, and \small$\PPositionVector{\InertialFrame}{\InertialFrame}{\PPropeller{p}} = \bm{\xi} +  \RotationMatrix{\InertialFrame}{\BodyFrame}  \PositionVector{\BodyFrame}{\BodyFrame}{\PPropeller{p}}$\normalsize, 
from which
\small\begin{align}
\VelocidadeAngular{\InertialFrame}{\InertialFrame}{\PPropeller{p}} &= \VelocidadeAngular{\InertialFrame}{\InertialFrame}{\BodyFrame}
= \RotationMatrix{\InertialFrame}{\BodyFrame} \VelocidadeAngular{\BodyFrame}{\InertialFrame}{\BodyFrame}
= \begin{bmatrix} \ZerosMatrix{3}{3} & \RotationMatrixInertialBody\JacobianoAngular_\eta\end{bmatrix}\dot{\bm{q}}
\triangleq \JacobianoAngular_{\PPropeller{p}}\dot{\bm{q}},
\label{AngPropeller} \\
\dPPositionVector{\InertialFrame}{\InertialFrame}{\PPropeller{p}} &= \VelocidadeLinear{\InertialFrame}{\InertialFrame}{\PPropeller{p}}
= \dot{\bm{\xi}} {-}  \RotationMatrix{\InertialFrame}{\BodyFrame} \SkewSymmetricMatrix{\PositionVector{\BodyFrame}{\BodyFrame}{\PPropeller{p}}}\JacobianoAngular_\eta\dot{\bm{\eta}}
= \begin{bmatrix} \IdentityMatrix{3} & -  \RotationMatrix{\InertialFrame}{\BodyFrame} \SkewSymmetricMatrix{\PositionVector{\BodyFrame}{\BodyFrame}{\PPropeller{p}}}\JacobianoAngular_\eta
\end{bmatrix}\dot{\bm{q}}
{\triangleq} \JacobianoLinear_{\PPropeller{p}}\dot{\bm{q}}. 
\label{LinPropeller}
\end{align} \normalsize
The thrust and torque generated by each propeller are given, respectively, by
\small\begin{equation}
    f_{\PPropeller{p}}(t) = b \Propeller{p}^2(t), \quad \tau_{\PPropeller{p}}(t) = \lambda_{\PPropeller{p}} k_\tau  \Propeller{p}^2(t),
    \label{ForceNTorquePropeller}
\end{equation} \normalsize
where \small$k_\tau \in \mathbb{R}$ \normalsize and \small$b \in \mathbb{R}$ \normalsize are the propeller's drag and thrust constants, \small$\Propeller{p}(t) \in \mathbb{R}$ \normalsize is propeller's angular velocity, and \small$\lambda_{\PPropeller{p}} \in \{1, -1\}$ \normalsize relates to the direction of rotation of the $p$-th propeller, in top view: if clockwise, \small$-1$\normalsize, if counter-clockwise, \small$1$\normalsize.

Taking into account \eqref{ForceNTorquePropeller}, \eqref{LinPropeller} and \eqref{AngPropeller}, the contribution of the $p$-th propeller to the vector of generalized coordinates is given by \small$\bm{\vartheta}_{\PPropeller{p}} = \JacobianoLinear'_{\PPropeller{p}} \RotationMatrix{\InertialFrame}{\PPropeller{p}}\bm{a}_z f_{\PPropeller{p}}(t) + \JacobianoAngular'_{\PPropeller{p}}\RotationMatrix{\InertialFrame}{\PPropeller{p}}\bm{a}_z \tau_{\PPropeller{p}}(t) = (\JacobianoLinear'_{\PPropeller{p}}  + \JacobianoAngular'_{\PPropeller{p}} \lambda_{\PPropeller{p}}   k_b) \RotationMatrix{\InertialFrame}{\PPropeller{p}}\bm{a}_z f_{\PPropeller{p}}(t)$\normalsize,  
where \small$\bm{a}_z \triangleq [0 \; 0 \; 1]'$, and $k_b \triangleq k_\tau/b$\normalsize. Therefore, the vector of generalized input is \small$\bm{\vartheta}(\bm{q},\bm{u}) = \sum^{8}_{p = 1} \bm{\vartheta}_{\PPropeller{p}}$\normalsize, yielding %
\small\begin{align}
    \bm{\vartheta}(\bm{q},\bm{u}) = \bm{B}(\bm{q})\bm{\tau} \triangleq \begin{bmatrix} \bm{\Xi}_1\bm{a}_z & \cdots & \bm{\Xi}_8\bm{a}_z    \end{bmatrix} \bm{\tau}, \quad \bm{\Xi}_i \triangleq \left(\JacobianoLinear'_{\PPropeller{i}}  {+} \JacobianoAngular'_{\PPropeller{i}} \lambda_{\PPropeller{i}}   k_b\right)\RotationMatrix{\InertialFrame}{\PPropeller{i}}. \label{eq:generalizedinputs}%
\end{align} \normalsize

\section{Nonlinear $\mathcal{W}_\infty$ control of the Octocopter UAV}
\label{Sec::Winf}

Before designing the nonlinear controller for the OUAV, the $\mathcal{W}_\infty$ control problem is stated considering a general second-order nonlinear nonautonomous dynamical system, and a particular solution to the resulting Hamilton-Jacobi (HJ) equation is proposed. Unlike in \cite{AUT2019}, this solution requires solving a single Riccati equation.

\begin{theorem}
\label{TheoremWinf}
Consider the nonlinear system 
\small\begin{gather}
\label{StCompactedSystem}
\textstyle\dot{\bm{\chi}}(t)=\bm{f}(\bm{\chi},t) + \bm{G}(\bm{\chi},t) \bm{u}(t) + \bm{K}(\bm{\chi},t)\bm{w}(t), \quad 
\bm{z}(t) = \int_{0}^{t}\tilde{\bm{q}}(\tau)d\tau,
\end{gather} \normalsize
with
    \small$\bm{f}(\bm{\chi},t) \triangleq \begin{bmatrix}
    \tilde{\bm{q}}'(t) & \dot{\tilde{\bm{q}}}'(t) & \bm{h}'(\bm{\chi},t)
    \end{bmatrix}',$
    $\bm{G}(\bm{\chi},t) = \bm{K}(\bm{\chi},t) = \big[
    \ZerosMatrix{}{} \,\; \ZerosMatrix{}{} \,\; \left(\bm{M}^{-1}(\bm{\chi},t)\right)'
    \big]'$\normalsize, %
where \small$\bm{u} \in \mathbb{R}^{n_u}$ \normalsize is the input vector, \small$\bm{w} \in \mathbb{R}^{n_u}$ \normalsize is the disturbance vector, \small$\tilde{\bm{q}} \triangleq \bm{q} - \bm{q}_r$\normalsize, where \small$\bm{q}(t) \in \mathbb{R}^{n_q}$ \normalsize are the generalized coordinates, \small$\bm{q}_r \in \mathcal{C}^2$ \normalsize is the desired reference, and \small$\bm{\chi}\triangleq \begin{bmatrix} \int_{0}^{\tau} \tilde{\bm{q}}'(\tau) d\tau \;\; \tilde{\bm{q}}'(t) \;\; \dot{\tilde{\bm{q}}}'(t)\end{bmatrix}'$ \normalsize is the state vector. Let \small$rank(\bm{M}(\bm{\chi},t)) = n_q$\normalsize, for any \small$\bm{\chi} \in \Omega$ \normalsize and \small$t \in \mathbb{R}_{\geq 0}$\normalsize, where \small$\Omega$ \normalsize stands for the set of all configurations the system can assume, with \small$n_q = n_u$\normalsize. Therefore, a control law that satisfies the nonlinear $\mathcal{W}_\infty$ optimal control (OC) problem %
\small\begin{align}
\label{StControlHinf}
\min_{\bm{u} \in \mathcal{U}}\max_{\bm{w} \in \mathcal{D}} & ~ \half ||\bm{z}(t)||^2_{\mathcal{W}_{3,2,\bm{\mathcal{Y}}}} - \half \gamma^2||\bm{w}(t)||_{\mathcal{L}_2}^2, &
&\text{s.t.}~~ \eqref{StCompactedSystem},
\end{align} \normalsize
for a given sufficiently large $\mathcal{W}_\infty$-index \small$\gamma \in \mathbb{R}_{\geq 0}$ \normalsize and weighting matrices \small$\bm{\mathcal{Y}} = \left(\bm{\mathcal{Y}}_0, \;  \bm{\mathcal{Y}}_1, \; \bm{\mathcal{Y}}_2, \;  \bm{\mathcal{Y}}_3 \right)$\normalsize, such that \small$\mathcal{U} = \mathbb{R}^{n_u}$ and $\mathcal{D} = \mathcal{L}_2[0,\infty)$\normalsize, is given by
\small\begin{gather}
\label{Stuopthinf}
\bm{u}(t) = -\bm{M}\left(\begin{bmatrix} \ZerosMatrix{}{} & \ZerosMatrix{}{} & (\bm{\mathcal{Y}}_3)^{-1} \end{bmatrix}\bm{Q}\bm{\chi} + \bm{h}(\bm{\chi},t)\right), ~~\text{s.t.}~~\bm{Q}\bm{A} {+} \bm{A}'\bm{Q} {-} \bm{Q}\bm{B}\bm{Q} {+} \bm{\Psi}{=} 0,
\end{gather}\normalsize
with \small$\bm{Q} > 0$\normalsize,
    \small$\bm{A} {\triangleq}
 \begin{bmatrix}
	\ZerosMatrix{}{} & \IdentityMatrix{} \\
	\ZerosMatrix{}{} & \ZerosMatrix{}{} 
	\end{bmatrix}, 
    \bm{B} {\triangleq} \text{blkdiag}(\ZerosMatrix{}{},\ZerosMatrix{}{},(\bm{\mathcal{Y}}_3)^{-1})$\normalsize,
  and \small$\bm{\Psi} \triangleq \text{blkdiag}(\bm{\mathcal{Y}}_0,\bm{\mathcal{Y}}_1,\bm{\mathcal{Y}}_2)$\normalsize.
\end{theorem}

\begin{proof}
By formulating the OC problem \eqref{StControlHinf} via dynamic programming, using differential game theory, the associated HJBI (Hamilton-Jacobi-Bellman-Isaacs) equation is given by%
\small\begin{align}
\label{hjifully}
\pderiv{\bm{V}_\infty}{t} + \min_{\bm{\tau} \in \mathcal{U}}\max_{\bm{w}_c \in \mathcal{D}}\{ \mathbb{H}_\infty(\bm{V}_\infty,\bm{\chi},\bm{u},\bm{w},t)\} = 0,
\end{align} \normalsize
with the Hamiltonian \small$\mathbb{H}_\infty \triangleq \pderivtrans{\bm{V}_{\infty}}{\bm{\chi}}\dot{\bm{\chi}} + \half (\bm{z}'\bm{\mathcal{Y}}_0\bm{z} + \dot{\bm{z}}'\bm{\mathcal{Y}}_1\dot{\bm{z}} + \ddot{\bm{z}}'\bm{\mathcal{Y}}_2\ddot{\bm{z}} + \dddot{\bm{z}}'\bm{\mathcal{Y}}_3\dddot{\bm{z}}-\gamma^2\bm{w}'\bm{w})$\normalsize, 
and boundary condition \small$\bm{V}_\infty(\ZerosMatrix{3m}{1},t) = 0$\normalsize. Considering $\bm{\Psi}$ as in the statement of the theorem, the Hamiltonian $\mathbb{H}_\infty$ is written in its expanded form as
\small\begin{align}
\label{HamiltinfFullyActuated}
\mathbb{H}_\infty &= \pderivtrans{\bm{V}_\infty}{\bm{\chi}} (\bm{f}(\bm{\chi},t) + \bm{G}(\bm{\chi},t) \bm{u} + \bm{K}(\bm{\chi},t)\bm{w})  
+ 
\half \bm{\chi}' \bm{\Psi} \bm{\chi}  \\  \nonumber
&+ \half \big[\bm{h}'\bm{\mathcal{Y}}_3\bm{h} + \bm{u}'(\bm{M}^{-1})'\bm{\mathcal{Y}}_3\bm{M}^{-1}\bm{u} + \bm{w}'(\bm{M}^{-1})'\bm{\mathcal{Y}}_3\bm{M}^{-1}\bm{w} + 2\bm{h}'\bm{\mathcal{Y}}_3\bm{M}^{-1}\bm{u} \\
&+ 2\bm{h}'\bm{\mathcal{Y}}_3\bm{M}^{-1}\bm{w} + 2\bm{u}'(\bm{M}^{-1})'\bm{\mathcal{Y}}_3\bm{M}^{-1}\bm{w}\big] - \half\gamma^2\bm{w}'\bm{w}\nonumber.
\end{align} \normalsize

The OC law, \small$\bm{u}^*$\normalsize, and the worst case of the disturbances, \small$\bm{w}^*$\normalsize, are obtained through the partial derivatives of \eqref{HamiltinfFullyActuated}, as follows:
\small\begin{align}
\label{D1FullyAct} 
\pderiv{\mathbb{H}_\infty}{\bm{u}} & = \bm{G}'\pderiv{\bm{V}_\infty}{\bm{\chi}}  + (\bm{M}^{-1})'\bm{\mathcal{Y}}_3\left(\bm{M}^{-1}\bm{u}^* + \bm{h} + \bm{M}^{-1}\bm{w}^*\right) = 0,\\
\label{D2FullyAct} 
\pderiv{\mathbb{H}_\infty}{\bm{w}} & = \bm{K}'\pderiv{\bm{V}_\infty}{\bm{\chi}} +(\bm{M}^{{-}1})'\bm{\mathcal{Y}}_3\left(\bm{M}^{-1}\bm{w}^* + \bm{h} + \bm{M}^{-1}\bm{u}^*\right) - \gamma^2\bm{w}^*= 0.
\end{align} \normalsize
Thus, considering \eqref{D1FullyAct}, the OC law is given by
\small\begin{align}
\nonumber
\bm{u}^* &= -\left((\bm{M}^{-1})'\bm{\mathcal{Y}}_3\bm{M}^{-1} \right)^{-1}\left(\bm{G}'\pderiv{\bm{V}_\infty}{\bm{\chi}}  + (\bm{M}^{-1})'\bm{\mathcal{Y}}_3\left(\bm{h} +\bm{M}^{-1}\bm{w}^*\right)\right), \\
 &= -\bm{M}\left(\begin{bmatrix} \ZerosMatrix{}{} & \ZerosMatrix{}{} & (\bm{\mathcal{Y}}_3)^{-1} \end{bmatrix}\pderiv{\bm{V}_\infty}{\bm{\chi}} + \bm{h}(\bm{\chi},t) + \bm{w}^*\right). \label{stuopthinf2}
\end{align} \normalsize
The worst case of the disturbances, $\bm{w}^*$, is computed by subtracting \eqref{D2FullyAct} from \eqref{D1FullyAct}, which yields
\vspace{-1mm}
\small\begin{gather}
\label{StWorstDist}
\bm{w}^* = (1/\gamma^2)\big( \bm{K}-\bm{G} \big)'\pderiv{\bm{V}_\infty}{\bm{\chi}}.
\end{gather} \normalsize
Through the second order partial derivatives of \eqref{D1FullyAct} and \eqref{D2FullyAct}, which are given, respectively, by
\small$\pderivsecond{\mathbb{H}_\infty}{\bm{u}} = (\bm{M}^{-1})'\bm{\mathcal{Y}}_3\bm{M}^{-1} > 0$\normalsize, and
\small$\pderivsecond{\mathbb{H}_\infty}{\bm{w}} = (\bm{M}^{-1})'\bm{\mathcal{Y}}_3\bm{M}^{-1}$ $- \gamma^2\bm{I} < 0$\normalsize,
it can be verified that \eqref{stuopthinf2} and \eqref{StWorstDist} are a Min-Max extremum of the OC problem for a sufficiently large \small$\gamma$\normalsize.

The HJ PDE associated to the problem is obtained by replacing the OC law \eqref{stuopthinf2}, and the worst case of the disturbances \eqref{StWorstDist}, in \eqref{hjifully}, yielding 
\small\begin{gather}
\label{StHinfProblem}
\pderiv{\bm{V}_\infty(\bm{\chi},t)}{t} + \mathbb{H}_{\infty}(\bm{V}_\infty,\bm{\chi},\bm{u}^*,\bm{w}^*,t) = 0.
\end{gather} \normalsize
Therefore, assuming %
	\small$\bm{V}_\infty(\bm{\chi}) = \half \bm{\chi}'\bm{Q} \bm{\chi} > 0$\normalsize,
	we have that \small$\pderiv{\bm{V}_\infty}{t} = 0$\normalsize. Consequently, \eqref{StHinfProblem} becomes
	\small\begin{align}
	\mathbb{H}_\infty(\bm{V}_\infty,\bm{\chi},\bm{u}^*,\bm{w}^*,t) = \pderivtrans{\bm{V}_\infty}{\bm{\chi}}\dot{\bm{\chi}} + 
	\half \bm{\chi}' \bm{\Psi} \bm{\chi} + \half \ddot{\tilde{\bm{q}}}'(\bm{\mathcal{Y}}_3)^{-1}\ddot{\tilde{\bm{q}}} = 0. \label{StHjb}
	\end{align} \normalsize
	
Taking into account \eqref{StCompactedSystem} and \eqref{StWorstDist}, we have that \small$\bm{w}^* = \ZerosMatrix{}{}$\normalsize, and from \eqref{stuopthinf2}, we have that
\small$\bm{u}^* = -\bm{M}\left(\begin{bmatrix} \ZerosMatrix{}{} & \ZerosMatrix{}{} & (\bm{\mathcal{Y}}_3)^{-1} \end{bmatrix}\bm{Q}\bm{\chi} + \bm{h}(\bm{\chi},t)\right)$\normalsize, %
which leads to
	\small$\ddot{\tilde{\bm{q}}} = \begin{bmatrix}
\ZerosMatrix{}{} & \ZerosMatrix{}{} & -(\bm{\mathcal{Y}}_3)^{-1}
\end{bmatrix}\bm{Q}\bm{\chi}$\normalsize, %
and the closed-loop (CL) state-space dynamics
	    \small$\dot{\bm{\chi}} = \begin{bmatrix}
	    \tilde{\bm{q}}' &
	    \dot{\tilde{\bm{q}}}' &
	    \ddot{\tilde{\bm{q}}}'
	    \end{bmatrix}' = %
        \mbf{A}\bm{\chi} 
	    + \mbf{B}%
     \bm{Q}\bm{\chi}$\small, 
 with \small$\mbf{A}$ \normalsize and \small$\mbf{B}$ \normalsize defined as in the statement of the theorem. By replacing these expressions for \small$\ddot{\tilde{\bm{q}}}$ and $\dot{\bm{\chi}}$ \normalsize
 in \eqref{StHjb}, we obtain the Riccati equation \small$\bm{Q}\bm{A} + \bm{A}'\bm{Q} - \bm{Q}\bm{B}\bm{Q} + \bm{\Psi}= 0$\normalsize, %
 which concludes the proof. \qed
	
\end{proof}

\begin{theorem}
\label{StabilityProofWinf}
    Let \small$\bm{q}_r \in \mathcal{C}^2$ \normalsize and \small$rank(\bm{M}(\bm{\chi},t)) = n_q$\normalsize, for any \small$\bm{\chi} \in \Omega$ \normalsize and \small$t \in \mathbb{R}_{\geq 0}$\normalsize. Therefore, the control law \eqref{Stuopthinf} provides asymptotic stability to system \eqref{StCompactedSystem}.%
\end{theorem}

\begin{proof}
Let $\check{\bm{\chi}} \triangleq (\bm{\chi}, \ddot{\bm{q}})$. As previously demonstrated, the CL system \eqref{StCompactedSystem} with the control law \eqref{Stuopthinf} ensures \eqref{StHjb} holds. Consequently, we have that %
\small%
	$\pderivtrans{\bm{V}_\infty}{\bm{\chi}}\dot{\bm{\chi}} = - 0.5 \check{\bm{\chi}}' \bm{\Theta} \check{\bm{\chi}},$ \normalsize
with $\bm{\Theta} \triangleq\text{blkdiag}( \bm{\Psi}, \bm{\mathcal{Y}}_3) > 0$, which means that \small$\bm{V}_\infty(\bm{\chi}) > 0$ \normalsize is a Lyapunov function that ensures asymptotic stability to the CL system. \qed
\end{proof}

To design the nonlinear $\mathcal{W}_\infty$ controller for the OUAV, system \eqref{EqCanonicaEulerLagrangeQuad} is partitioned into controlled, \small$\bm{q}_c$\normalsize, and regulated DOF, \small$\bm{q}_r$\normalsize,
\small\begin{align}
\label{ContUncontSystem}
\begin{bmatrix}
	\bm{M}_{cc} & \bm{M}_{cr} \\\bm{M}_{rc} & \bm{M}_{rr}
	\end{bmatrix} \ddot{\mbf{q}}(t) {+} \begin{bmatrix}
	\bm{C}_{cc} & \bm{C}_{cr} \\\bm{C}_{rc} & \bm{C}_{rr}
	\end{bmatrix}
\dot{\mbf{q}}(t){+}\begin{bmatrix}
	\bm{g}_c \\
	\bm{g}_r
	\end{bmatrix} {=} \begin{bmatrix}
	\bm{B}_c \\
	\bm{B}_r
	\end{bmatrix}\bm{\tau} {+} \mbf{w}(t),
\end{align} \normalsize
where \small$\bm{q}(t) \in \mathbb{R}^{6}$\normalsize, with \small$\bm{q}(t) = [\bm{q}'_c(t) \;\bm{q}'_r(t)]'$\normalsize, in which \small$\bm{q}_c(t) \triangleq [x(t) \; y(t) \; z(t) ]'$ \normalsize corresponds to the controlled DOF and \small$\bm{q}_r(t) \triangleq [\phi(t) \; \theta(t) \; \psi(t)]'$ \normalsize to the regulated DOF, and $\bm{w}(t) = \begin{bmatrix}
	\bm{w}'_c &
	\bm{w}'_r
	\end{bmatrix}'$. %
Then, from  \eqref{ContUncontSystem} and considering \small$\tilde{\bm{q}}_c = \bm{q}_c - \bm{q}_{c_r}$\normalsize, we take into account the tracking error dynamics of the controlled DOF, %
\small\begin{align}
\label{ContUncontSystem3}
	\ddot{\tilde{\bm{q}}}_c %
	 &= \bm{M}^{-1}_{cc} \left( - \bm{M}_{cr} \ddot{\bm{q}}_r - \bm{C}_{cc}\dot{\bm{q}}_c - \bm{C}_{cr}\dot{\bm{q}}_r
	-\bm{g}_c \right)+ \bm{M}^{-1}_{cc}\bm{u} + \bm{M}^{-1}_{cc}\bm{w}_c - \ddot{\bm{q}}_{c_r},
\end{align} \normalsize
with \small$\bm{u} \triangleq \bm{B}_c\bm{\tau}$\normalsize, where \small$\bm{q}_{c_r} \in \mathcal{C}^2$ \normalsize stands for the desired reference set to the controlled DOF. Finally, by defining the state vector \small$\bm{\chi}\triangleq [\int_{0}^{\tau} \tilde{\bm{q}}'_c d\tau \,\; \tilde{\bm{q}}'_c \,\; \dot{\tilde{\bm{q}}}'_c]'$\normalsize, %
system \eqref{ContUncontSystem3} is represented in the standard form %
\small\begin{equation}
\label{CompactedSystem1}
\textstyle\dot{\bm{\chi}}(t)=\bm{f}(\bm{q},\dot{\bm{q}},\ddot{\bm{q}}_r,t) + \bm{G}(\bm{q}) \bm{u} + \bm{K}(\bm{q})\bm{w}_r, \quad \bm{z}_c(t) = \int_{0}^{t}\tilde{\bm{q}}_c(\tau)d\tau,
\end{equation} \normalsize
with \small$\bm{z}_c(t)$ \normalsize being the cost variable selected as an integral action over the tracking error \small$\tilde{\bm{q}}_c$ \normalsize to provide parametric uncertainty and constant disturbance rejection capability for the CL system, and
\small$\bm{f}(\bm{q},\dot{\bm{q}},\ddot{\bm{q}}_r,t) \triangleq \begin{bmatrix}
    \tilde{\bm{q}}'_c & \dot{\tilde{\bm{q}}}'_c & \bm{h}'_c
    \end{bmatrix}$, with $\bm{h}_c \triangleq \bm{M}^{-1}_{cc} ( - \bm{M}_{cr} \ddot{\bm{q}}_r - \bm{C}_{cc}\dot{\bm{q}}_c - \bm{C}_{cr}\dot{\bm{q}}_r	-\bm{g}_c )- \ddot{\bm{q}}_{c_r}$\normalsize, and \small$\bm{K}(\bm{q}) = \bm{G}(\bm{q}) \triangleq \begin{bmatrix}
    \ZerosMatrix{}{} & \ZerosMatrix{}{} & \bm{M}^{-1}_{cc}
    \end{bmatrix}'$\normalsize. 

Considering \eqref{CompactedSystem1} and Theorem \ref{TheoremWinf}, the nonlinear $\mathcal{W}_\infty$ control law \eqref{Stuopthinf} is obtained, which is given in terms of the generalized input \small$\bm{u}^* = \bm{B}_c\bm{\tau}$\normalsize. From \eqref{eq:generalizedinputs}, 
\small\begin{align}
\bm{u}^* &= \bm{B}_c\bm{\tau} = \RotationMatrix{\InertialFrame}{\BodyFrame}\bm{a}_z f_z
= \begin{bmatrix}
\cos(\psi) & -\sin(\psi)& 0 \\
\sin(\psi)& \cos(\psi)& 0 \\
        0&        0& 1
\end{bmatrix}\begin{bmatrix}
         \cos(\phi) \sin(\theta) \\
         -\sin(\phi)\\
         \cos(\phi) \cos(\theta)
         \end{bmatrix}f_z, \label{ControlInputsAllocation1}
\end{align} \normalsize
where \small$\bm{B}_c =  \begin{bmatrix}
    \RotationMatrix{\InertialFrame}{\BodyFrame}\bm{a}_z & \RotationMatrix{\InertialFrame}{\BodyFrame}\bm{a}_z & \cdots & \RotationMatrix{\InertialFrame}{\BodyFrame}\bm{a}_z
    \end{bmatrix}$\normalsize, and %
\small$f_z \triangleq f_{\PPropeller{1}} + f_{\PPropeller{2}} + \cdots + f_{\PPropeller{8}}$ \normalsize is the total thrust. Therefore, to satisfy \eqref{ControlInputsAllocation1}, we use the following control allocation scheme:
\small\begin{align}
\label{ControlAllocation1} 
\textstyle
    f_{z} = {u_3}/\left(\cos(\phi) \cos(\theta)\right), 
    \phi_r = \text{asin}\left(-{u_2}/{f_{z}}\right), 
    \theta_r = \text{asin}\left({u_1}/\big({\cos(\phi_r)f_{z}}\big)\right),
\end{align} \normalsize
for \small$f_{z}>0$\normalsize, \small$\cos(\phi) \neq 0$\normalsize, and \small$\cos(\theta) \neq 0$\normalsize, where \small$\bm{u}_i$\normalsize, for \small$i \in \{1, 2, 3\}$\normalsize, stands to the element corresponding to the i-th row of vector \small$\bm{u}^*$\normalsize, and \small$\phi_r$ \normalsize and \small$\theta_r$ \normalsize stands for the desired references for the roll and pitch angles.

To comply with \eqref{ControlAllocation1}, a second nonlinear $\mathcal{W}_\infty$ controller is designed taking into account the dynamics of the regulated DOF. Therefore, by replacing \small$\ddot{\bm{q}}_c$ \normalsize in the second row of \eqref{ContUncontSystem},
the tracking error dynamics of \small$\bm{q}_r$ \normalsize are written as %
\small\begin{gather}
\label{sys2}
\ddot{\tilde{{\bm{q}}}}_r = -\bar{\bm{M}}^{-1}\big(\bar{\bm{C}}\dot{\tilde{\bm{q}}}_r +\bar{\bm{M}}\ddot{\bm{q}}_{r_r} + \bar{\bm{C}}\dot{\bm{q}}_{r_r}+ \bar{\bm{e}}\big) + \bar{\bm{B}}\bm{\tau} + \bar{\bm{\delta}},
\end{gather} \normalsize
where \small$\tilde{\bm{q}}_r \triangleq \bm{q}_r - \bm{q}_{r_r}$\normalsize, in which \small$\bm{q}_{r_r} \in \mathcal{C}^{2}$ \normalsize stands for the desired value of the regulated DOF, with \small$\phi_r$ \normalsize and \small$\theta_r$ \normalsize given in \eqref{ControlAllocation1}. Besides, \small$\bar{\bm{M}}(\bm{q}) \triangleq \bm{M}_{rr} - \bm{M}_{rc}\bm{M}^{-1}_{cc}\bm{M}_{cr}$, $\bar{\bm{C}}(\bm{q},\dot{\bm{q}}) \triangleq \bm{C}_{rr} - \bm{M}_{rc}\bm{M}^{-1}_{cc}\bm{C}_{cr}$\normalsize, \small$\bar{\bm{e}}(\bm{q},\dot{\bm{q}}) \triangleq \bm{g}_r - \bm{M}_{rc}\bm{M}^{-1}_{cc}\bm{g}_{c} + \big(\bm{C}_{rc} - \bm{M}_{rc}\bm{M}^{-1}_{cc}\bm{C}_{cc}\big)\dot{\bm{q}}_c$\normalsize, \small$\bar{\bm{B}}(\bm{q}) = \bm{B}_r - \bm{M}_{rc}\bm{M}^{-1}_{cc}\bm{B}_{c}$\normalsize, and \small$\bar{\bm{\delta}}(\bm{q},t) \triangleq \bm{w}_r - \bm{M}_{rc}\bm{M}^{-1}_{cc}\bm{w}_c$\normalsize. %

\begin{remark}
    In this work, we set \small$\dot{\phi}_r =\dot{\theta}_r = 0$\normalsize, for control design purposes based on the time scale separation hypothesis of a cascade control strategy.
\end{remark}
Given the state vector \small$\bm{x} \triangleq \big[ \int_0^{t}\tilde{\bm{q}}'_rd\tau \; \tilde{\bm{q}}'_r \; \dot{\tilde{\bm{q}}}'_r
\big]'$\normalsize, \eqref{sys2} is represented as
\small\begin{gather}
\label{CompactedSystemFullyActuatedHinf} 
\textstyle\dot{\bm{x}}= \bar{\bm{f}}(\bm{q},\dot{\bm{q}},t) + \bar{\bm{g}}(\bm{q},t)\bar{\bm{u}}+\bar{\bm{k}}(\bm{q},t)\bar{\bm{\delta}}, \quad \bm{z}_r = \int_{0}^{t}\tilde{\bm{q}}_c(\tau)d\tau,
\end{gather} \normalsize
with \small$\bm{z}_r$ \normalsize being the cost variable, where \small$\bar{\bm{u}} \triangleq \bar{\bm{B}}\bm{\tau}$\normalsize, \small$\bar{\bm{d}} \triangleq \bar{\bm{M}}\ddot{\bm{q}}_{r_r} + \bar{\bm{C}}\dot{\bm{q}}_{r_r} + \bar{\bm{e}}$\normalsize, and %
\small$\bar{\bm{f}}(\bm{q},\dot{\bm{q}},t) \triangleq \big[ \tilde{\bm{q}}'_r \; \dot{\tilde{\bm{q}}}'_r \; h_r \big]'$\normalsize, \small$\bar{\bm{k}}(\bm{q}) = \bar{\bm{g}}(\bm{q}) \triangleq \begin{bmatrix}
  \ZerosMatrix{m}{m}& \ZerosMatrix{m}{m} & \bar{\bm{M}}^{-1}
 \end{bmatrix}'$\normalsize. 
The nonlinear $\mathcal{W}_\infty$ control law \eqref{Stuopthinf} is obtained taking into account \eqref{CompactedSystem1} and Theorem \ref{TheoremWinf}.
This OC law is given in terms of the generalized input, \small$\bar{\bm{u}}^* = \bar{\bm{B}}\bm{\tau}$\normalsize. To obtain \small$\bm{\tau}$\normalsize, we propose the following control allocation scheme:
\small\begin{align}
	\label{ControlAllocation}
	\min_{\bar{\bm{\tau}}\in\mathbb{R}^{8}}& ~\half\left(\bar{\bm{u}}^* - \bar{\bm{B}}\bm{\tau}\right)'\left(\bar{\bm{u}}^* - \bar{\bm{B}}\bm{\tau}\right), ~~~~\text{s.t.}~~~~~\text{i), ~~ii), ~~\text{and}~~ iii)},
\end{align} \normalsize
with i) \small$\bar{\bm{u}}^* = -\bar{\bm{M}}\big([ \ZerosMatrix{}{} \,\; \ZerosMatrix{}{} \,\; (\bm{\mathcal{Y}}_3)^{-1} ] \bm{Q}\bm{x} -\bar{\bm{M}}^{-1}\bar{\bm{C}}\dot{\tilde{\bm{q}}}_r -\bar{\bm{M}}^{-1}\bar{\bm{d}}\big)$\normalsize, ii) \small$f_{\PPropeller{i}} - f_{\PPropeller{i+1}} = 0$ \normalsize and iii) \small$\sum_{i=1}^8f_{\PPropeller{i}} = f_z,~ \forall i \in \{1, 3, 5, 7\}$\normalsize.

\begin{remark}
    The minimum $\bar{\bm{u}}^* =  \bar{\bm{B}}\bm{\tau}$ to \eqref{ControlAllocation} constitutes a set of equations that can be satisfied even with the imposition of constraints ii) and iii) (i.e. five constraints), since $\bm{\tau} \in \mathbb{R}^{8}$ and $\bar{\bm{u}}^* \in \mathbb{R}^{3}$.
\end{remark}

\section{Joint state, input and parameter estimation}%
\label{Sec:Estimator}

\vspace{-1mm}
Consider the nonlinear discrete-time system
\small\begin{equation} \label{eq:systemUKF}
\begin{aligned}
\mbf{x}_{k} & = \bm{\varphi}(\mbf{x}_{k-1}, \mbf{u}_{k-1},\mbf{d}_{k-1},\mbf{p}_{k-1}) + \mbf{B}_w \mbf{w}_{k-1}, \\
\mbf{y}_{k} & = \bm{\pi}(\mbf{x}_{k}, \mbf{u}_{k},\mbf{d}_k,\mbf{p}_{k}) + \mbf{D}_v \mbf{v}_{k}, \\
\end{aligned}
\end{equation} \normalsize
where \small$\mbf{x}_k \in \realset^{n}$ \normalsize is the system state, \small$\mbf{u}_{k} \in \realset^{n_u}$ \normalsize is the known input, \small$\mbf{w}_k \in \realset^{n_w}$ \normalsize is the process disturbance, \small$\mbf{y}_k \in \realset^{n_y}$ \normalsize is the measured output, \small$\mbf{v}_k \in \realset^{n_v}$ \normalsize is the measurement disturbance, \small$\mbf{d}_k \in \realset^{n_d}$ \normalsize are unknown exogenous inputs, and \small$\mbf{p}_k \in \realset^{n_p}$ \normalsize are the unknown parameters. In addition, \small$\bm{\varphi}: \realset^n \times \realset^{n_u} \times \realset^{n_d} \times \realset^{n_p} \to \realset^{n}$ \normalsize and \small$\bm{\pi}: \realset^n \times \realset^{n_u} \times \realset^{n_d} \times \realset^{n_p} \to \realset^{n_y}$ \normalsize are nonlinear mappings. The initial state, exogenous input, model parameters, and disturbances are assumed to satisfy \small$\mbf{x}_0 \sim \mathcal{N}(\hat{\mbf{x}}_0, \mbf{P}^{x}_0)$, $\mbf{d}_0 \sim \mathcal{N}(\hat{\mbf{d}}_0, \mbf{P}^{d}_0)$\normalsize, \small$\mbf{d}_k - \mbf{d}_{k-1} \triangleq \Delta \mbf{d}_{k-1} \sim \mathcal{N}(\mbf{0}, \mbf{P}^{\Delta d})$, $\mbf{p}_0 \sim \mathcal{N}(\hat{\mbf{p}}_0, \mbf{P}^{p}_0)$\normalsize, \small$\mbf{p}_k - \mbf{p}_{k-1} \triangleq \Delta \mbf{p}_{k-1} \sim \mathcal{N}(\mbf{0}, \mbf{P}^{\Delta p})$, $\mbf{w}_k \sim \mathcal{N}(\mbf{0}, \mbf{P}^{w})$\normalsize, and \small$\mbf{v}_k \sim \mathcal{N}(\mbf{0}, \mbf{P}^{v})$\small. %
In order to estimate the exogenous input \small$\mbf{d}_k$ \normalsize and the unknown parameters \small$\mbf{p}_k$\normalsize, we consider the augmented state vector \small$\bm{\mu}_k \triangleq (\mbf{x}_k, \mbf{d}_k, \mbf{p}_k) \in \realset^{n_\mu}$\normalsize, with \small$n_\mu \triangleq n+n_d+n_p$\normalsize, \small$\bm{\mu}_k \sim \mathcal{N}(\hat{\bm{\mu}}_0, \mbf{P}^\mu_0)$\normalsize, where \small$\hat{\bm{\mu}}_0 \triangleq (\hat{\mbf{x}}_0, \hat{\mbf{d}}_0, \hat{\mbf{p}}_0)$\normalsize, and \small$\mbf{P}_0^\mu \triangleq \text{blkdiag}(\mbf{P}^{x}_0, \mbf{P}^{d}_0, \mbf{P}^{p}_0)$\normalsize. The system \eqref{eq:systemUKF} is then rewritten in terms of the augmented variable \small$\bm{\mu}_k$ \normalsize as
\small\begin{equation} \label{eq:systemUKFjoint}
\bm{\mu}_{k} = \tilde{\bm{\varphi}}(\bm{\mu}_{k-1}, \mbf{u}_{k-1}) + \tilde{\mbf{B}}_w \tilde{\mbf{w}}_{k-1}, \quad \mbf{y}_{k} = \tilde{\bm{\pi}}(\bm{\mu}_{k}, \mbf{u}_{k}) + \mbf{D}_v \mbf{v}_{k}
\end{equation} \normalsize
where \small$\tilde{\bm{\varphi}}: \realset^{n_\mu} \times \realset^{n_u} \to \realset^{n_\mu}$, $\tilde{\bm{\pi}}: \realset^{n_\mu} \times \realset^{n_u} \to \realset^{n_y}$\normalsize, \small$\tilde{\mbf{B}}_w \in \realsetmat{n_\mu}{n_w+n_d+n_p}$ and $\tilde{\mbf{w}}_{k-1} \in \realset^{n_w+n_d+n_p}$ \normalsize are given by \small$\tilde{\bm{\varphi}}(\bm{\mu}_{k-1},\mbf{u}_{k-1}) \triangleq (\bm{\varphi}(\mbf{x}_{k-1},\mbf{u}_{k-1},\mbf{d}_{k-1},\mbf{p}_{k-1}), $ $\mbf{d}_{k-1}, \mbf{p}_{k-1})$\normalsize, \small$\tilde{\bm{\pi}}(\bm{\mu}_{k}, \mbf{u}_{k}) \triangleq \bm{\pi}(\mbf{x}_{k}, \mbf{u}_{k},\mbf{d}_k,\mbf{p}_{k})$\normalsize, \small$\tilde{\mbf{B}}_w \triangleq \text{blkdiag}(\mbf{B}_w, \eye{n_d}, \eye{n_p})$\normalsize, and \small$\tilde{\mbf{w}}_{k-1} \triangleq (\mbf{w}_{k-1}, \Delta \mbf{d}_{k-1}, \Delta \mbf{p}_{k-1})$\normalsize. %

Given the augmented system \eqref{eq:systemUKFjoint}, \small$\bm{\mu}_0 \sim \mathcal{N}(\hat{\bm{\mu}}_0, \mbf{P}^{\mu}_0)$, $\tilde{\mbf{w}}_{k-1} \sim \mathcal{N}(\mbf{0}, \mbf{P}^{\tilde{w}})$\normalsize, \small$\mbf{v}_{k} \sim \mathcal{N}(\mbf{0}, \mbf{P}^{v})$\normalsize, with \small$\mbf{P}^{\tilde{w}} \triangleq \text{blkdiag}(\mbf{P}^{w}, \mbf{P}^{\Delta d}, \mbf{P}^{\Delta p})$\normalsize, the objective is to estimate the tuple \small$(\hat{\bm{\mu}}_k, \hat{\mbf{P}}^\mu_k)$ \normalsize for known input and measurement sequences \small$(\mbf{u}_0, \mbf{u}_1, \ldots, \mbf{u}_{k-1})$ and $(\mbf{y}_1, \mbf{y}_2, \ldots, \mbf{y}_{k})$\normalsize, respectively. This is done recursively through the joint state, input, and parameter unscented Kalman filter (JUKF) composed of the prediction and update steps described in the following paragraphs.%

Let \small$\bm{\mu}_{k-1} \sim \mathcal{N}(\hat{\bm{\mu}}_{k-1}, \hat{\mbf{P}}^\mu_{k-1})$\normalsize, and consider the \textit{sigma-point transformation} \citep{Julier2000} \small$[\hat{\bm{\mu}}^{(1)}_{k-1} \,\; \hat{\bm{\mu}}^{(2)}_{k-1} \,\; \cdots \,\; \hat{\bm{\mu}}^{(2n)}_{k-1}] \triangleq \hat{\bm{\mu}}_{k-1} \ones{1}{2n} + \sqrt{n} [\hat{\mbf{S}}^\mu_{k-1} \,\; {-}\hat{\mbf{S}}^\mu_{k-1}]$\normalsize, %
where \small$\hat{\bm{\mu}}^{(i)}_{k-1}$ \normalsize is the $i$-th sigma point associated to \small$\mathcal{N}(\hat{\bm{\mu}}_{k-1}, \hat{\mbf{P}}^\mu_{k-1})$\normalsize, and \small$\hat{\mbf{S}}^\mu_{k-1}$ \normalsize is the lower triangular matrix obtained from the Cholesky decomposition \small$\hat{\mbf{S}}^\mu_{k-1} (\hat{\mbf{S}}^\mu_{k-1})^T = \hat{\mbf{P}}^\mu_{k-1}$\normalsize. %
Let \small$\hat{\bm{\mu}}^{(i)+}_{k-1} \triangleq \tilde{\bm{\varphi}}(\hat{\bm{\mu}}^{(i)}_{k-1}, \mbf{u}_{k-1})$\normalsize, and \small$\rho \triangleq 1/(2n)$\normalsize, then the predicted tuple \small$(\bar{\bm{\mu}}_k, \bar{\mbf{P}}^\mu_{k})$ \normalsize is obtained by the \emph{prediction step} \small$\bar{\bm{\mu}}_k = \rho \sum_{i=1}^{2n}\hat{\bm{\mu}}^{(i)+}_{k-1}$\normalsize, \small$\bar{\mbf{P}}^\mu_{k} = \tilde{\mbf{B}}_w \mbf{P}^{\tilde{w}} \tilde{\mbf{B}}_w^T +  \rho \sum_{i=1}^{2n}(\hat{\bm{\mu}}^{(i)+}_{k-1} - \bar{\bm{\mu}}_k)(\hat{\bm{\mu}}^{(i)+}_{k-1} - \bar{\bm{\mu}}_k)^T$\normalsize. 
Moreover, consider the sigma-point transformation \small$[\bar{\bm{\mu}}^{(1)}_{k} \,\; \bar{\bm{\mu}}^{(2)}_{k} \,\; \cdots \,\; \bar{\bm{\mu}}^{(2n)}_{k}] \triangleq \bar{\bm{\mu}}_{k} \ones{1}{2n} + \sqrt{n} \left[\bar{\mbf{S}}^\mu_{k} \,\; {-}\bar{\mbf{S}}^\mu_{k}\right]$\normalsize, where \small$\bar{\mbf{S}}^\mu_{k}$ \normalsize is the lower triangular matrix obtained from the Cholesky decomposition \small$\bar{\mbf{S}}^\mu_{k} (\bar{\mbf{S}}^\mu_{k})^T = \bar{\mbf{P}}^\mu_{k}$\normalsize. 
Define \small$\bar{\mbf{y}}^{(i)+}_k \triangleq \tilde{\bm{\pi}}(\bar{\bm{\mu}}^{(i)}_{k}, \mbf{u}_{k})$\normalsize, then the predicted tuple \small$(\bar{\mbf{y}}_k, \bar{\mbf{P}}^y_{k}, \bar{\mbf{P}}^{\mu y}_{k})$ \normalsize is obtained by \small$\bar{\mbf{y}}_k = \rho \sum_{i=1}^{2n}\bar{\mbf{y}}^{(i)+}_k, \quad \bar{\mbf{P}}^y_{k} = \mbf{D}_v \mbf{P}^v (\mbf{D}_v)^T $ $+  \rho \sum_{i=1}^{2n}(\bar{\mbf{y}}^{(i)+}_{k} - \bar{\mbf{y}}_k)(\bar{\mbf{y}}^{(i)+}_{k} - \bar{\mbf{y}}_k)^T$\normalsize, \small$\bar{\mbf{P}}^{\mu y}_{k} = \rho \sum_{i=1}^{2n}(\bar{\bm{\mu}}^{(i)}_{k} - \bar{\bm{\mu}}_k)(\bar{\mbf{y}}^{(i)+}_{k} - \bar{\mbf{y}}_k)^T$\normalsize. 

Finally, consider the tuple \small$(\bar{\bm{\mu}}_k, \bar{\mbf{P}}^\mu_{k}, \bar{\mbf{y}}_k, \bar{\mbf{P}}^y_{k}, \bar{\mbf{P}}^{\mu y}_{k})$ \normalsize obtained in the prediction step. The Kalman gain \small$\mbf{K} \in \realsetmat{n_\mu}{n_y}$ \normalsize is computed by \small$\mbf{K}_k = \bar{\mbf{P}}^{\mu y}_{k} (\bar{\mbf{P}}^{y}_{k})^{-1}$\normalsize, 
with the estimated tuple \small$(\hat{\bm{\mu}}_k,\hat{\mbf{P}}_k^{\mu})$ \normalsize being obtained by the \emph{update step} \small$\hat{\bm{\mu}}_k = \bar{\bm{\mu}}_k + \mbf{K}_k (\mbf{y}_k - \bar{\mbf{y}}_k)$, $\hat{\mbf{P}}^\mu_k = \bar{\mbf{P}}^\mu_k - \mbf{K}_k  \bar{\mbf{P}}^y_k \mbf{K}_k^T$\normalsize.

\subsection{Load parameterization, sensors, and control structure} \label{sec:JUKFparameterization}

For the sake of simplicity, we consider that the load has a cubic shape with edge length $2 r_\text{L}$, and it is assumed to have homogeneous mass distribution. By defining $\mbf{p}_k \triangleq (m_{\FrameC{L}}, r_\text{L}) \in \realset^{2}$, %
the inertia tensor \small$\mathbb{I}_{\FrameC{L}}$ \normalsize and the displacement vector \small$\PositionVector{\BodyFrame}{\BodyFrame}{\FrameC{L}}$ \normalsize are then parameterized as \small$\mathbb{I}_{\FrameC{L}}(\mbf{p}_k) \triangleq (m_{\FrameC{L}}/6)r_\text{L}^2 \eye{3}$\normalsize, and \small$\PositionVector{\BodyFrame}{\BodyFrame}{\FrameC{L}}(\mbf{p}_k) \triangleq (0, 0, \varsigma (\delta_\text{L} + r_\text{L}))$\normalsize, respectively, 
where \small$\varsigma = -1$ \normalsize if the load is attached below the UAV, \small$\varsigma = 1$ \normalsize if it is attached above the UAV, and \small$\delta_\text{L}$ \normalsize is a known displacement from the origin of $\mathcal{B}$ to the point of attachment of the load.

Consider \eqref{EqCanonicaEulerLagrangeQuad}, with \small$\bm{\vartheta}(\mbf{q}(t),\bm{\tau}(t)) = \mbf{B}(\mbf{q}(t))\bm{\tau}(t)$ \normalsize as in \eqref{eq:generalizedinputs}. Let \small$\mbf{x}(t) \triangleq (\mbf{q}(t), \dot{\mbf{q}}(t)) \in \realset^{12}$ \normalsize and \small$\mbf{u}(t) \triangleq \bm{\tau}(t) \in \realset^8$\normalsize, and using the forward Euler discretization method (FEDM) with sampling time \small$T_s$\normalsize, we have that \small$\mbf{x}_k =\mbf{x}_{k-1} + T_s \left.\dot{\mbf{x}}(t)\right|_{t = (k-1)T_s} + \bm{\ell}_{k-1}$\normalsize, where \small$\bm{\ell}_{k-1} \in \realset^{12}$ \normalsize is the discretization error. %
Moreover, for state estimation, we consider that %
\small$\bm{\zeta}_k \triangleq \begin{bmatrix} \eye{2} & \zeros{2}{4} \end{bmatrix}' \mbf{d}_k +  \bm{\varpi}_k$\normalsize, 
with \small$\mbf{d}_k \in \realset^2$ \normalsize being exogenous inputs, and \small$\bm{\varpi}_k \in \realset^{6}$ \normalsize being modeling uncertainties. In addition, we define the vector of process disturbances as \small$\mbf{w}_{k-1} \triangleq \begin{bmatrix} \zeros{6}{6} \\ (\mbf{M}(\mbf{x}_{k-1},\mbf{p}_{k-1})^{-1} \end{bmatrix}\bm{\varpi}_{k-1} + \bm{\ell}_{k-1}$\normalsize. %
Hence, the state-space equations for joint estimation of the OUAV are given in discrete time by \small$\mbf{x}_k = \bm{\varphi}(\mbf{x}_{k-1},\mbf{u}_{k-1},\mbf{d}_{k-1},\mbf{p}_{k-1}) + \mbf{w}_{k-1}$\normalsize, where
\small\begin{align}
\bm{\varphi}(\mbf{x}_{k-1},\mbf{u}_{k-1},\mbf{d}_{k-1},\mbf{p}_{k-1}) \triangleq \mbf{x}_{k-1} + T_s \begin{bmatrix} [\mbf{0} \,\; \eye{6}] \mbf{x}_{k-1} \\ \bm{\iota}(\mbf{x}_{k-1},\mbf{u}_{k-1},\mbf{d}_{k-1},\mbf{p}_{k-1})\end{bmatrix}, \label{eq:filter_f} 
\end{align}\normalsize
and \small$\bm{\iota}(\mbf{x}_{k-1},\mbf{u}_{k-1},\mbf{d}_{k-1},\mbf{p}_{k-1}) \triangleq (\mbf{M}(\mbf{x}_{k-1},\mbf{p}_{k-1}))^{-1}(- \mbf{C}(\mbf{x}_{k-1},\mbf{p}_{k-1}) [\mbf{0} \,\; \eye{6}] \mbf{x}_{k-1} - \mbf{g}(\mbf{x}_{k-1},\mbf{p}_{k-1}) + \mbf{B}(\mbf{x}_{k-1})\mbf{u}_{k-1} + \begin{bmatrix} \eye{2} & \zeros{2}{4} \end{bmatrix}' \mbf{d}_{k-1}\Big)$\normalsize. 

For design of the JUKF, we consider that a Global Positioning System (GPS), a barometer, an accelerometer, a gyroscope and a magnetometer are available onboard the OUAV. The following variables are assumed to be measured: the position of the UAV w.r.t to the inertial frame $\mathcal{I}$, \small$\bm{\xi}_k = [x_k\,\; y_k \,\; z_k]^T$\normalsize, the orientation of the UAV w.r.t. $\mathcal{I}$, \small$\bm{\eta}_k = [\phi_k \,\; \theta_k \,\; \psi_k]^T$\normalsize, and the angular velocity \small$\VelocidadeAngular{\mathcal{B}}{\mathcal{I}}{\mathcal{B}}$\normalsize. The measurement \small$\mbf{y}_k \in \realset^9$ \normalsize is described by the measurement equation \small$\mbf{y}_{k} = \bm{\pi}(\mbf{x}_{k}, \mbf{u}_{k},\mbf{d}_k,\mbf{p}_{k}) + \mbf{v}_{k}$\normalsize, with \small$\bm{\pi}(\mbf{x}_{k}, \mbf{u}_{k},\mbf{d}_k,\mbf{p}_{k}) \triangleq (\bm{\xi}_k, \bm{\eta}_k, \VelocidadeAngular{\mathcal{B}}{\mathcal{I}}{\mathcal{B}}) = (\bm{\xi}_k, \bm{\eta}_k,$ $\JacobianoAngular_\eta(\bm{\eta}_k) \dot{\bm{\eta}}_k)$\normalsize, 
where \small$\JacobianoAngular_\eta(\bm{\eta}_k)$ \normalsize defined as in Section \ref{OctocopterUAVmodel}. %
Finally, let \small$\bm{\mu}_k \triangleq (\mbf{x}_k, \mbf{d}_k, \mbf{p}_k) \in \realset^{16}$\normalsize, with \small$\bm{\mu}_k \sim \mathcal{N}(\hat{\bm{\mu}}_0, \mbf{P}^{\mu}_0)$\normalsize, where \small$\hat{\bm{\mu}}_0 \triangleq (\hat{\mbf{x}}_0, \hat{\mbf{d}}_0, \hat{\mbf{p}}_0)$\normalsize, and \small$\mbf{P}_0 \triangleq \text{blkdiag}(\mbf{P}^{x}_0, \mbf{P}^{d}_0, \mbf{P}^{p}_0)$\normalsize. Then, \small$\tilde{\bm{\varphi}}$\normalsize, \small$\tilde{\bm{\pi}}$\normalsize, \small$\tilde{\mbf{B}}_w$\normalsize, \small$\mbf{D}_v$ \normalsize in \eqref{eq:systemUKFjoint} are \small$\tilde{\bm{\varphi}}(\bm{\mu}_{k-1}$, $ \mbf{u}_{k-1}) \triangleq (\bm{\varphi}(\mbf{x}_{k-1},$ $\mbf{u}_{k-1},\mbf{d}_{k-1},$ $\mbf{p}_{k-1}), \mbf{d}_{k-1}, \mbf{p}_{k-1})$\normalsize, \small$\tilde{\bm{\pi}}(\bm{\mu}_{k}, \mbf{u}_{k}) \triangleq (\bm{\xi}_k, \bm{\eta}_k, $ $\JacobianoAngular_\eta(\bm{\eta}_k) \dot{\bm{\eta}}_k)$\normalsize, \small$\tilde{\mbf{B}}_w = \eye{16}$\normalsize, \small$\mbf{D}_v = \eye{9}$\normalsize, with \small$\bm{\varphi}(\mbf{x}_{k-1},\mbf{u}_{k-1},\mbf{d}_{k-1},\mbf{p}_{k-1})$ \normalsize defined as in \eqref{eq:filter_f}. %
Hence, the JUKF for the OUAV with load is composed by the prediction and update steps described previously in this section. %
Using the known input \small$\mbf{u}_k$ \normalsize and the measurement vector \small$\mbf{y}_k$\normalsize, the JUKF provides estimates of the states and load parameters to the nonlinear $\mathcal{W}_\infty$ controller.%

\section{Numerical experiments}

\vspace{-1mm}
This section conducts a numerical experiment, using MATLAB R2017a, to corroborate the efficacy of the proposed control strategy. The differential equations composing the control system were discretized using the FEDM with sampling time \small$T_s = 0.01$ \normalsize s, and the OUAV physical parameters are given in Table \ref{OctoParamaters} (drag and thrust constants estimated based on the brushless TMotor U15II KV100 with propeller G40x13.1CF, \url{https://store.tmotor.com/goods.php?id=732}). %
Moreover, for simulation purposes, \small$m_{\FrameC{L}} = 100$ \normalsize kg and \small$r_\text{L} = 0.5$ \normalsize m. 

\begin{figure}[!tb]
	\centering{
	\def\svgwidth{0.98\columnwidth}
	{\tiny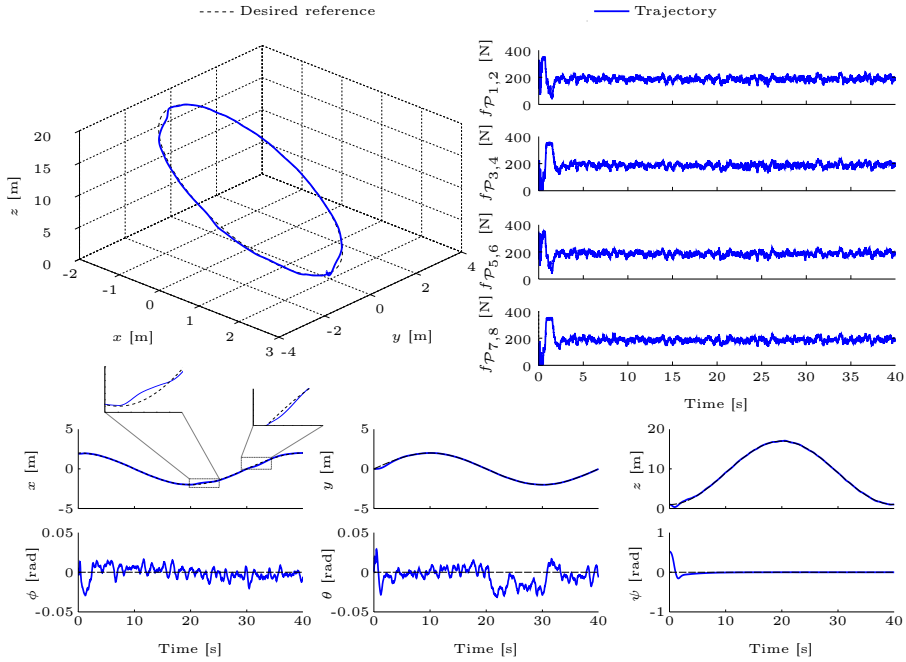}
	\caption{The OUAV trajectory, and the time evolution of $\mbf{u}(t)$ and $\mbf{q}(t)$.}\label{FigureResults1}}
\end{figure}

\begin{figure}[tb!]
	\centering{
	\def\svgwidth{0.9\columnwidth}
    {\tiny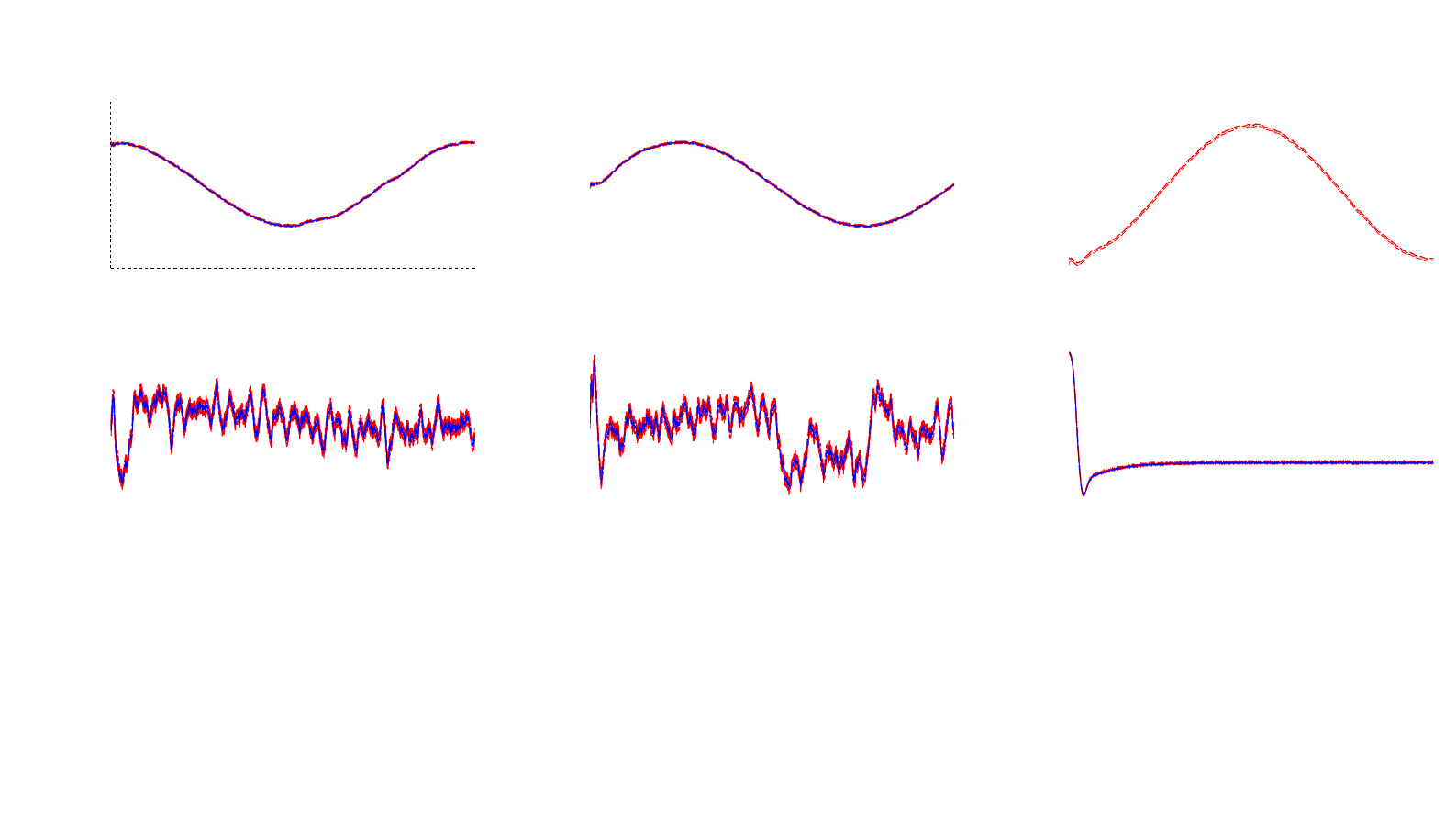}
	\caption{Estimated generalized DOF, disturbances and load parameters.}	\label{FigureResults2}}
\end{figure}

To implement the nonlinear $\mathcal{W}_\infty$ controller proposed in Section \ref{Sec::Winf}, the OC law, \small$\bm{u}^*$\normalsize, for the controlled DOF dynamics \eqref{CompactedSystem1}, was tuned via Brynson's method. Then, a fine tuning was performed over the initial tuning, which resulted in 
\small$\mbf{\mathcal{Y}}^{(c)}_0 = \text{diag}(5, 5, 10)$\normalsize,
\small$\mbf{\mathcal{Y}}^{(c)}_1 = \text{diag}(10, 10, 50)$\normalsize,
\small$\mbf{\mathcal{Y}}^{(c)}_2 = \text{diag}(1, 1, 1)$\normalsize,
\small$\mbf{\mathcal{Y}}^{(c)}_3 = \text{diag}(6, 6, 1)$\normalsize.
The same tuning process was considered to the OC law, \small$\bm{u}^*$\normalsize, for the regulated DOF dynamics \eqref{CompactedSystemFullyActuatedHinf}, which resulted in
\small$\mbf{\mathcal{Y}}^{(r)}_0 = \text{diag}(1, 1, 1)$\normalsize,
\small$\mbf{\mathcal{Y}}^{(r)}_1 = \text{diag}(10, 10, 10)$\normalsize,
\small$\mbf{\mathcal{Y}}^{(r)}_2 = \text{diag}(0.2, 0.2, 0.2)$\normalsize,
\small$\mbf{\mathcal{Y}}^{(r)}_3 = \text{diag}(0.05, 0.05, 0.05)$\normalsize.
To execute the nonlinear $\mathcal{W}_\infty$ controller, first, the desired values to the roll, \small$\phi_r$\normalsize, and pitch, \small$\theta_r$\normalsize, angles, and the total thrust \small$f_{z}$ \normalsize are computed from \eqref{ControlAllocation1}. Then, the OC allocation scheme \eqref{ControlAllocation} is solved to obtain $\bm{\tau}$, which is applied to \eqref{EqCanonicaEulerLagrangeQuad}. The nonlinear $\mathcal{W}_\infty$ controller is computed using $(\hat{\mbf{x}}_k,\hat{\mbf{p}}_k)$ estimated by the JUKF proposed in Section \ref{Sec:Estimator}. The controller is executed at each sampling time $T_s$, while the discrete-time design of the JUKF takes advantage of the sampling nature of the measurement $\mbf{y}_k$. %
The JUKF was designed with \small$\mbf{P}^{\tilde{w}} = \text{diag}((\frac{0.01}{3})^2 {\cdot} \ones{12}{1}, (\frac{1}{3})^2 {\cdot} \ones{2}{1},(\frac{2}{3})^2, $ $(\frac{0.001}{3})^2)$, $\mbf{P}^v = \text{diag}(0.05^2 {\cdot} \ones{2}{1}, \;0.17^2, \;(\frac{0.05\pi}{180})^2{\cdot} \ones{3}{1}, \;0.00552^2{\cdot} \ones{3}{1})$\normalsize, \small$\mbf{P}^x_0 =  \text{diag}((\frac{2}{3})^2 {\cdot} \ones{3}{1}, $ $(\frac{\pi}{18})^2 {\cdot} \ones{2}{1}, (\frac{\pi}{6})^2, (\frac{1}{3}) {\cdot} \ones{3}{1}, (\frac{\pi}{36})^2 {\cdot} \ones{3}{1})$\normalsize, \small$\mbf{P}_0^{d} {=} \text{diag}((\frac{1}{3})^2 {\cdot} \ones{2}{1})$\normalsize, \small$\mbf{P}_0^{p} {=} \text{diag}((\frac{50}{3})^2,(\frac{0.75}{3})^2)$, $\bar{\mbf{x}}_0 {=} (2.4,0.5,-0.2,\frac{\pi}{6},\frac{\pi}{6},0,\zeros{6}{1})$\normalsize, \small$\bar{\mbf{d}}_0 = (0,0)$, and $\bar{\mbf{p}}_0 = (50,0.75)$\normalsize. 
The OUAV was prompted to perform the desired trajectory \small$x_r(t) =2\cos(2\pi t/40)$ \normalsize m, \small$y_r(t) = 2\sin(2 \pi t/40)$ \normalsize m, \small$z_r(t) = 9 - 8\cos(2\pi t/40)$ \normalsize m, and \small$\psi_r(t) = 0$ \normalsize rad, starting displaced from the desired trajectory with the initial conditions \small$\bm{q}(0) = [1.9\; 0\; 0.8\; 0\; 0\; \pi/6]'$ \normalsize and \small$\dot{\bm{q}}(0) = \ZerosMatrix{0}{0}$\normalsize. The disturbance \small$\bm{\zeta}_x(t) = 30$ \normalsize N, for \small $ 20 \leq t \leq 30$\normalsize, is applied to the system. Figure \ref{FigureResults1} presents the results obtained by the control strategy, and Figure \ref{FigureResults2} illustrates the time evolution of the states, exogenous disturbances, and load parameters estimated by the JUKF. %
It can be noticed that the JUKF was able to estimate the generalized DOF, \small$\mbf{q}$\normalsize, the generalized velocities, \small$\dot{\mbf{q}}$\normalsize, the disturbances, \small$\bm{\zeta}_x$ \normalsize and \small$\bm{\zeta}_y$\normalsize, and the load parameters, \small$m_{\FrameC{L}}$ \normalsize and \small$r_\text{L}$\normalsize, and provided these information to the nonlinear $\mathcal{W}_\infty$ controller. The proposed robust control scheme based on the multi-body approach was capable of performing the heavy load transportation subjected to load parameter uncertainties, while maintaining the system stable and attenuating the effects of the exogenous disturbances. %

\begin{table}[!tb]
	\centering
 	\setlength\tabcolsep{1pt}
	\renewcommand{\arraystretch}{0.3}
	\caption{Octocopter UAV physical parameters.}
	\label{OctoParamaters}
    \footnotesize
	\begin{tabular}{cccc}
		\hline
		Parameter & Value & Parameter & Value\\\hline
		$\mathbb{I}_{\FrameC{O}}$	&  diag$(18.78, 19.76, 37.87) $ kg.m$^2$ & $\alpha_1, \alpha_2$ & $\pi/4$ $(45)$  rad (deg) \\
		$m_{\FrameC{O}}$         &  $53.09$ kg & $\alpha_3, \alpha_4$ & $3\pi/4$ $(135)$  rad (deg)\\
  		$g$         &  $-9.81$ m/s$^2$ & $\alpha_5, \alpha_6$ & $5\pi/4$ $(225)$ rad (deg)\\	
		$\lambda_{\PPropeller{1}}, \lambda_{\PPropeller{3}}, \lambda_{\PPropeller{5}}, \lambda_{\PPropeller{7}}$ & 1 & $\alpha_7, \alpha_8$ & $7\pi/4$ $(315)$ rad (deg) \\
		$\lambda_{\PPropeller{2}}, \lambda_{\PPropeller{4}}, \lambda_{\PPropeller{6}}, \lambda_{\PPropeller{8}}$ & -1 & $b$			& $2.85\cdot10^{-5}$ N${\cdot}$s$^2$ \\
		$\PositionVector{\BodyFrame}{\BodyFrame}{\FrameC{O}}$ & $(0,0,0)$ m & $k_\tau$         & $1.42\cdot10^{-6}$ N${\cdot}$m${\cdot}$s$^2$\\
		$\PositionVector{\BodyFrame}{\BodyFrame}{\PPropeller{1}}$ & $(1.1,1.1,0.12)$ m & $\PositionVector{\BodyFrame}{\BodyFrame}{\PPropeller{5}}$ & $({-}1.1,{-}1.1,0.12)$ m \\
		$\PositionVector{\BodyFrame}{\BodyFrame}{\PPropeller{2}}$ & $(1.1,1.1,{-}0.17)$ m & $\PositionVector{\BodyFrame}{\BodyFrame}{\PPropeller{6}}$ & $({-}1.1,{-}1.1,{-}0.17)$ m \\
		$\PositionVector{\BodyFrame}{\BodyFrame}{\PPropeller{3}}$ & $({-}1.1,1.1,0.12)$ m & $\PositionVector{\BodyFrame}{\BodyFrame}{\PPropeller{7}}$ & $(1.1,{-}1.1,0.12)$ m \\
		$\PositionVector{\BodyFrame}{\BodyFrame}{\PPropeller{4}}$ & $({-}1.1,1.1,{-}0.17)$ m & 		$\PositionVector{\BodyFrame}{\BodyFrame}{\PPropeller{8}}$ & $(1.1,{-}1.1,{-}0.17)$ m \\
  		$m_{\FrameC{L}}$         &  $\in [0,100]$ kg & $\mathbb{I}_{\FrameC{L}}$	&  $(m_{\FrameC{L}}/6)r_\text{L}^2 {\cdot} \eye{3}$ kg.m$^2$\\
        $r_\text{L}$             & $\in [0,1.5]$ m & $\PositionVector{\BodyFrame}{\BodyFrame}{\FrameC{L}}$ & $(0,0,- \varsigma (\delta_\text{L} + r_\text{L}))$ m \\  
		\hline
	\end{tabular} \normalsize %
\end{table}


\vspace{-1mm}
\section{Conclusions}

This work proposed a joint state-parameter observer-based controller for trajectory tracking of an OUAV, for transportation of a heavy load with unknown mass and size. A multi-body dynamic model of the OUAV with a rigidly attached load was obtained, effectively considering the effects of the load parameters. A nonlinear $\mathcal{W}_\infty$ controller was designed for optimal trajectory tracking of the OUAV, with states and load parameters provided by a UKF-based algorithm. The effectiveness of our approach was corroborated by numerical results, which was capable of performing the load transportation subjected to parameter uncertainties and exogenous disturbances. 
Future work will consider a wind disturbance model, the load connected through a rope, and perform real flight experiments.

\vspace{-1mm}
\small
\bibliography{References,RegoMThesis_bib}

\begin{thebibliography}{10}

\bibitem{Bernard2011}
M.~Bernard, K.~Kondak, I.~Maza, and A.~Ollero.
\newblock Autonomous transportation and deployment with aerial robots for
  search and rescue missions.
\newblock {\em Journal of Field Robotics}, 28(6):914--931, October 2011.

\bibitem{BisgaardThesis}
M.~Bisgaard.
\newblock {\em Modeling, Estimation and Control of Helicopter Slung Load
  System}.
\newblock PhD thesis, Aalborg University, 2008.

\bibitem{Bisgaard2010}
M.~Bisgaard, A.~l. Cour-Harbo, and J.~D. Bendtsen.
\newblock Adaptive control system for autonomous helicopter slung load
  operations.
\newblock {\em Control Engineering Practice}, 18(7):800--811, July 2010.

\bibitem{AUT2019}
D.~N. Cardoso, S.~R. Esteban, and G.~V. Raffo.
\newblock A robust optimal control approach in the weighted sobolev space for
  underactuated mechanical systems.
\newblock {\em Automatica}, 125:1--11, 2021.

\bibitem{Gajbhiye2022}
S.~Gajbhiye, D.~Cabecinhas, C.~Silvestre, and R.~Cunha.
\newblock Geometric finite-time inner-outer loop trajectory tracking control
  strategy for quadrotor slung-load transportation.
\newblock {\em Nonlinear Dynamics}, 107(3):2291--2308, 2022.

\bibitem{Julier2000}
S.~Juiler, J.~Uhlmann, and H.~F. Durrant-Whyte.
\newblock A new method for the nonlinear transformations of means and
  covariances in filters and estimations.
\newblock {\em IEEE Trans. on Automatic Control}, 45(3):477--482, 2000.

\bibitem{Palunko2013}
I.~Palunko, A.~Faust, P.~Cruz, L.~Tapia, and R.~Fierro.
\newblock A reinforcement learning approach towards autonomous suspended load
  manipulation using aerial robots.
\newblock In {\em Proc. of the IEEE ICRA}, pages 4881--4886, May 2013.

\bibitem{Prkacin2020}
V.~Prka$\check{\text{c}}$in, I.~Palunko, and I.~Petrovi\'{c}.
\newblock State and parameter estimation of suspended load using quadrotor
  onboard sensors.
\newblock In {\em Proc. of the IEEE 2020 ICUAS}, pages 958--967, 2020.

\bibitem{spong2006robot}
M.~W. Spong, S.~Hutchinson, M.~Vidyasagar, et~al.
\newblock {\em Robot modeling and control}, volume~3.
\newblock Wiley New York, 2006.

\bibitem{Sreenath2013b}
K.~Sreenath, T.~Lee, and V.~Kumar.
\newblock Geometric control and differential flatness of a quadrotor {UAV} with
  a cable-suspended load.
\newblock In {\em Proc. of the 52nd IEEE CDC}, pages 2269--2274, December 2013.

\bibitem{Wang2014}
F.~Wang, P.~Liu, S.~Zhao, B.~M. Chen, S.~K. Phang, S.~Lai, T.~H. Lee, and
  C.~Cai.
\newblock Guidance, navigation and control of an unmanned helicopter for
  automatic cargo transportation.
\newblock In {\em Proc. of the IEEE 33rd CCC}, pages 1013--1020, July 2014.

\bibitem{Yu2022}
G.~Yu, D.~Cabecinhas, R.~Cunha, and C.~Silvestre.
\newblock Adaptive control with unknown mass estimation for a
  quadrotor-slung-load system.
\newblock {\em ISA Transactions}, 133:412--423, 2023.

\end{thebibliography}
\normalsize

\end{document}